# RAMSES: A full-stack application for detecting seizures and reducing data during continuous EEG monitoring

John M. Bernabei, Olaoluwa Owoputi, Shyon D. Small, Nathaniel T. Nyema, Elom Dumenyo, Joongwon Kim, Steven N. Baldassano, Christopher Painter, Erin C. Conrad, Taneeta M. Ganguly, Ramani Balu, Kathryn A. Davis, Jay Pathmanathan, Brian Litt

*Abstract*—*Objective:* Continuous EEG (cEEG) monitoring is associated with lower mortality in critically ill patients, however it is underutilized due to the difficulty of manually interpreting prolonged streams of cEEG data. Here we present a novel real-time, machine learning-based alerting and monitoring system for epilepsy and seizures (RAMSES) that dramatically reduces the amount of manual EEG review. *Methods:* We developed a custom data reduction algorithm using a random forest, and deployed it within an online cloud-based platform which streams data and communicates interactively with caregivers via a web interface to display algorithm results. We validate RAMSES on cEEG recordings from 77 patients undergoing routine scalp ICU EEG monitoring. *Results:* On subjects with seizures we achieved >80% overall data reduction, while detecting a mean of 84% of seizures across all validation patients, with 19/27 patients achieving 100% seizure detection. On seizure free-patients, the majority of cEEG records, we reduced data requiring manual review by >83%. *Conclusion:* This study validates a platform for machine-learning assisted data reduction. *Significance:* This work represents a meaningful step towards improving utility and decreasing cost for cEEG monitoring We also make our high-quality annotated dataset of 77 ICU cEEG recordings public for others to validate and improve upon our methods.

Manuscript received August 10, 2020. J. M. Bernabei acknowledges support from NIH 6T32NS091006. B. Litt acknowledges support from the Penn Center for Healthcare Innovation and the Mirowski Family Foundation. K. A. Davis acknowledges support from NIH K23-NS073801 and the Thornton Foundation. (Corresponding author: John M. Bernabei)

J. M. Bernabei, O. Owoputi, S. D. Small, N. T. Nyema, E. Dumenyo, J. Kim, S. N. Baldassano, and C. R. Painter are with the Department of Bioengineering and the Center for Neuroengineering & Therapeutics, University of Pennsylvania, Philadelphia, PA 19104 USA (e-mail: johnbe@seas.upenn.edu, owo@seas.upenn.edu, shyons@seas.upenn.edu, nyema@seas.upenn.edu, edumenyo@seas.upenn.edu, jkim0118@seas.upenn.edu, stevennb@pennmedicine.upenn.edu, chrispainter96@gmail.com).

E. C. Conrad, T. M. Ganguly, R. Balu, K. A. Davis, J. Pathmanathan, and B. Litt are with the Department of Neurology, University of Pennsylvania, Philadelphia, PA 19104 USA (e-mail: erin.conrad2@uphs.upenn.edu, taneeta.ganguly@uphs.upenn.edu, ramani.balu@pennmedicine.upenn.edu, katedavis@uphs.upenn.edu, jay.pathmanathan@pennmedicine.upenn.edu, littb@upenn.edu).

## I. INTRODUCTION

CONTINUOUS EEG (cEEG) monitoring is often performed in the intensive care unit to assess cerebral activity in real time[1] and detect seizures, which are associated with worse patient outcomes[2], [3]. Continuous EEG is the test of choice for detecting non – convulsive status epilepticus (NCSE), which is commonly found in ICU patients and carries a high mortality rate[4]. Unfortunately, prompt recognition of NCSE is challenging. At present there are no fully automated systems for monitoring cEEG. Data review requires manual interpretation by trained experts such as physicians or EEG technologists[5]. In addition, demand for cEEG monitoring often fluctuates widely, (in our hospital system ranging from 1 to over 20 cEEG patients monitored at any given time), making staffing challenging and costly[6]. Even when such staff are in place, they are often limited in the number of patients they are able to monitor due to labor-intensive demands for maintaining the quality of scalp recordings and rapidly interpreting recordings. At many institutions cEEG is often read at scheduled intervals, usually 8 to 12 hours, which can lead to significant delays in identification of critical events. Automated systems for evaluating cEEG in real time have the potential to recognize actionable events, such as seizures, much more quickly than manual interpretation, and at a lower cost per patient, which could expand the use of cEEG in both resource-rich and resource-poor healthcare settings[7].

Visual quantitative EEG (qEEG) methods have been deployed to reduce the time and cost associated with manual EEG interpretation[8], [9]. The most commonly used qEEG techniques offer near real-time analysis, displaying compressed metrics derived from amplitude and frequency. Changes in these parameters can be used to detect seizures and cortical ischemia. However, while qEEG may significantly reduce review time by the clinician, sensitivity for identifying seizures remains low (51-67%)[8]. Despite its advantages over inspection of raw waveform data, visual qEEG still requires specialized training and the inspection of the entire recording albeit in a compressed format.

As an alternative to visual qEEG, there are several algorithms for detecting seizures in scalp EEG. The most widespread is Persyst's "Reveal" algorithm, which has a reported clinical sensitivity of 76% with a false positive rate of 0.11/Hour[10], though subsequent studies have shown a significantly higher false-positive rate[11]. This level of performance leads to a significant proportion of seizures being missed, as well as a high false alarm burden. Other "non-patient-specific" algorithms have been reported to perform better than Persyst, but many previous studies use epilepsy patients known to have stereotyped seizures[12] or test on a carefully curated and cleaned dataset. Patient-specific algorithms have the highest level of performance [13] but require clinicians to mark training data for each individual, which, depending on the time to the first event, renders these approaches less practical for deploying rapidly in an ICU. There is a clear clinical need for non-patient-specific seizure detection algorithms that are highly sensitive and specific on the noisy, artifact-heavy, and heterogeneous data typical of the ICU. There is also a need for a gold standard, widely available cEEG data set and objective performance criteria for seizure detection algorithms that can be used by experts and the FDA for benchmarking cEEG analysis tools, similar to what our group has done for benchmarking seizure detection algorithms for intracranial EEG[14].

In this study we introduce a novel framework for semi – automated cEEG analysis and data reduction developed using data collected in the intensive care unit. We share our source code and unique dataset openly for others to improve upon our results and methods. Rather than designing an algorithm to replace clinical cEEG review entirely, we use machine learning to perform data reduction with the intent of increasing the speed and decreasing the cost required to accurately evaluate cEEG for seizures. Furthermore, our framework includes a data streaming portal providing simplified yet detailed data to expedite treatment decisions or guide further EEG review. Overall, we aim to establish a path for our methods to be easily translated into clinical care, regardless of EEG hardware, in a way that can quickly scale cEEG monitoring in hospitals worldwide.

## II. Materials and Methods

Our retrospective dataset consists of 77 patients who were treated in the ICU at hospitals in the University of Pennsylvania Health System and underwent cEEG monitoring between 2017 and 2019. We used 27 randomly selected records with discrete seizures, and 50 consecutive records without seizures after excluding separate records belonging to the same patient. Data were collected in concordance with the institutional review board of the University of Pennsylvania. EEG signals were recorded and digitized at 256 Hz using Natus Xltek equipment (Natus Medical, Pleasanton CA). All EEGs were acquired using a 10-20 electrode configuration with lateral eye leads at minimum. EEG recordings were annotated by board certified clinical neurophysiologists to include times of onset and offset for all seizures. We stored EEG recordings on http://ieeg.org[15], a cloud platform for storing and sharing electrophysiologic data.

### A. Feature Extraction

We filtered raw EEG signals using a 5th order Bessel bandpass filter with cutoff frequencies of 1 Hz and 20 Hz and calculated features using a non-overlapping sliding 5 second window. We calculated the following features for each channel in each window and used the median value of each feature across channels: (i) Power in the delta (1 - 4 Hz), theta (4 - 8 Hz), alpha (8 - 12 Hz), and beta (12 - 25 Hz) frequency bands, (ii) signal line length[16], which quantifies the distance between successive points and has been shown to be an effective feature in seizure detection, (iii) wavelet entropy[17], which measures the signal complexity in the time and frequency domains and has proven to be an effective EEG feature[18], (iv) statistical features including mean, variance, and kurtosis, and (vi) the mean value of the upper signal envelope of the EEG waveform. Within each window we used an automated artifact rejection algorithm to remove channels containing missing values or supraphysiologic amplitudes that were clearly due to noise, and also excluded any 5 second window with at least three channels containing missing values or shared artifacts that would introduce error into algorithm. At the beginning of feature calculation during model training, the artifact rejection algorithm begins with conservative threshold values of each feature and iteratively rejects segments that surpass those feature levels, and checks whether any of the rejected segments were clinically labeled seizures. If so, the threshold of each feature for artifact rejection is raised 50% and the process is repeated, yielding criteria which will not incorrectly reject seizure as artifact in any training patients.

### B. Machine learning approach

We implemented a machine learning framework that identifies EEG segments of high seizure likelihood in unseen patients. Our algorithmic approach is summarized in Figure 1. We use clinically annotated seizures in which the unequivocal seizure onset (UEO) and offset on EEG are marked by board-certified EEG readers using the method of Litt et al.[19] (Figure 1A), and calculate the mean and variance of each feature (Figure 1B) across all channels from all time windows in this initial segment, yielding a total of 20 features. We train a random forest classifier (Figure 1C) using 400 trees to predict whether each 5 second EEG segment in each patient contains a seizure or not. As cEEG segments that contain seizures make up only a small proportion of our overall dataset, reflecting clinical practice, we train the classifier to be penalized 500 times as heavily for false negatives as for false positives. After calculating and normalizing patient features, we used five-fold cross-validation in which 1/5 of the patients were held out of training at a time to be used for validation. After the model generates its prediction for each window, the prediction is then briefly post-processed. We fill in any gaps of a single window in length that are marked by the system as not being seizure to smooth predictions (Figure 1D). Specifically, we use a filter size of 4 windows to in-fill seizure predictions between nearby windows, and subsequently remove seizure predictions from single 5-second windows that are not within 15 seconds of other identified seizure segments as these are unlikely to contain a seizure. This step makes the reduced EEG significantly more contiguous and amenable to clinical review.

### C. Evaluation

To measure the performance of our model, we calculated 'seizure sensitivity' and 'data reduction'. We define 'seizure sensitivity' as the proportion of seizures for which the algorithm marks either the entirety or a portion of the event on EEG. We define "data reduction" as the proportion of data removed from consideration by the detection system. We calculate this metric by determining the proportion of true negative windows in the patient's time series that are marked for removal rather than review by the clinician. MATLAB's default random forest function can output class probabilities, allowing us to generate receiver-operating characteristic curves to assess the tradeoff between seizure sensitivity and data reduction by varying the detection (default = 0.5). Each clinician user can adjust this threshold to tune the seizure detection – data reduction tradeoff to their own preferences.

### D. Full-Stack Application

We construct an integrated application to manage the inputs, outputs, data storage of our novel machine learning algorithm, and its interaction with users. In our application, an open-source task management platform called "Celery" allocates separate, asynchronous processes which harvest and process data from EEG streams, calculate seizure likelihood using the algorithm, and ultimately store predictions in a MongoDB database (Figure 2A). In this study we used http://ieeg.org on Amazon's elastic computing cloud for EEG storage and its toolboxes for data streaming into MATLAB, but implementation could be performed on local machines behind institutional firewalls, or on HIPAA compliant cloud facilities, as optimal for individual institutions. We also present an interactive, web platform using the Python Flask library to display reduced EEG and allow clinicians to interact with and understand the outputs of our system (Figures 2B, 2C).

The main page of the dashboard shows an overview of all patients who are currently undergoing cEEG monitoring in the ICU with the RAMSES system. Each patient is listed with information related to outputs from the classifier. Specifically, we show the number of seizures detected over the course of a patient's recording, the percentage of the recording consisting of concerning epochs, the time in minutes since the last seizure, and a visual representation of the most concerning prediction over the length of the recording. This visual representation appears as a dot color coded according to the respective prediction as follows: red is a likely seizure, yellow is a potential seizure and blue is non seizure. Clinicians may also adapt the exact dashboard layout and display statistics according to their preferences. Specifically, they may order patients according to either the most recent seizures, the highest density of concerning epochs or the room number. Furthermore, they may adjust the time period over which the statistics are calculated to provide different quantifications of clinical status.

Within the same application, clinicians are also afforded the option to select any given patient and further inspect the algorithm's outputs for them over the duration of the recording. In this patient-specific view, predictions are represented as a timeline with different epochs color coded in accordance with the color associations of the dots on the main page. Clinicians can further inspect the raw EEG associated with each prediction by double-clicking the prediction on timeline.

### E. Data and Code Sharing

The patient dataset used to develop our algorithm represents one of the largest published machine learning studies of annotated ICU cEEG. All records and annotations are freely available on http://ieeg.org in the project ICU_Monitoring. The code for the seizure detection and data reduction algorithms is available at GitHub.com/jbernabei/ICU_EEG, while the code for streaming and web-interfacing is available at GitHub.com/nathanielnyema/RAMSES. We aim for our methods to be translatable and for other groups to validate and improve our algorithms or their own using the resources that we provide.

### III. RESULTS

We retrospectively acquired data from 77 critical care patient cEEG recordings including individuals with and without seizures. cEEG records were clinically read in their entirety and annotated by boarded epileptologists for the presence of seizures, and their onset and offset times were marked. Patient metadata collected for this study is included in Supplementary Table 1. To assess seizure detection performance, and its trade off with the amount of data reduction, we calculated cross validation seizure sensitivity across patients. Figure 3A shows the ROC curve representing the performance of our model. At an 80% data reduction (equal to specificity) we achieve a seizure sensitivity of mean 84%, median 100%. Figure 3B shows the distribution of model performance across patients. Of the 27 patients who had seizures during the recording period, the data reduction algorithm only missed all seizures in a one patient, who had a single 10-second event lacking high-frequency activity over a low-voltage background (Supplementary Figure 1). In all other patients few seizures were missed. At this seizure detection sensitivity, the algorithm achieved 83% data reduction (specificity) in the 50 seizure-free patients evaluated. Figure 3A shows the tradeoff between seizure detection sensitivity and the amount of data reduction achieved by our algorithms.

To provide a better understanding of the strengths and limitations of a clinical implementation of our system, we visualize examples of system outputs in Figure 4. Panel A shows true-positive and false-positive EEG examples for a patient in which the algorithm correctly identified 13/18 seizures while achieving a 99.2% data reduction. On the left we show a clip of correctly classified seizure activity localized to the right hemisphere while on the right we show a non-seizure segment which the algorithm erroneously classified as possibly containing a seizure. The asymmetry of right and left hemisphere activity in the false-positive example could have skewed our algorithm to predict seizure during this time segment. Panel B shows a patient in which our algorithm correctly identified 36/40 seizures with a 91.7% data reduction. On the left, a correctly identified seizure is displayed, while on the right there is an example of a missed seizure where strong discharges in the frontal electrodes may have masked the high frequency activity in feature space. This algorithm achieved a seizure detection sensitivity of 90% for this patient.

## IV. Discussion

In this study we present an important step in developing and implementing automated cEEG analysis systems to manage the increasing demand for expensive ICU EEG monitoring. Our main objective was to use machine learning to both reliably detect seizures and to dramatically reduce the amount of EEG that must be physically reviewed by physicians and trained technologists. Additionally, we aimed to provide an open source framework to allow data handling, storage, and display – which could be applied to other uses of EEG monitoring (for example, vasospasm and ischemia monitoring, prognostication, and assessment of level of sedation). We found that this approach provides >80% data reduction when tuned to broad detection performance across patients (mean seizure detection of 84%, median of 100% of seizures across patients). Our full-stack application provides functionality for data streaming as well as displaying results in a customizable web dashboard. We share all of our data and code with the intention that our methods are improved upon so that machine-learning assisted data reduction can be used clinically to expand the use and decrease the cost of continuous EEG in the ICU setting.

A key question in studies of this nature is what analysis performance metrics are adequate for clinical deployment. Seizure sensitivity is the typical gold standard, though seizure labeling varies significantly between experts[20]. Furthermore, the open nature of our processing algorithm allows our algorithm to train towards any given experts reporting style. There is also some indication that to adequately manage patients, it may not be necessary to capture and identify every seizure on EEG, as many of these events may not have clinical significance, and trends in number of events, combined with clinical metadata may be adequate for excellent patient management even if a small percentage of subtle electrographic seizures are missed by the algorithm. While there are no available studies of rates of missed seizures when humans continuously review huge streams of patient data, either from multiple patients simultaneously in real time, as is done in institutions and private agencies that provide centralized monitoring, our experience is that the percentage of missed seizures is likely similar to our algorithm performance.

Many promising automated EEG analysis tools have failed to reach clinical use despite promising initial studies. One reason for this may be challenges of integrating them into a clinical workflow. For example, many algorithms were designed to alleviate the need of clinicians to review EEG entirely and attempt to precisely identify seizures and send clinical alerts. However, without large clinical trials and FDA approval, it would be difficult for these algorithms to be implemented to replace typical EEG review. We propose that data reduction tool such as ours may be more amenable to clinical use, as it does not aim to replace clinical judgement and could be easily tuned for the seizure sensitivity and data reduction tradeoff each user prefers. Another reason for the limited adoption of other machine learning models for EEG analysis is that many are developed on private datasets which might not be representative of the vast heterogeneity and noise present in ICU cEEG recordings at a large academic center. Usually these data sets are not made available for community access and validation, as software developers for clinically deployed systems usually work on proprietary commercial systems. Finally, most published algorithms may not have the input and display functionality necessary for clinical deployment. In these cases, it is impossible for these algorithms to be implemented without commercializing the technology into a costly, licensed product. To circumvent these problems, we have made all of our EEG records and code publicly available so that potential users can inspect the characteristics and quality of our underlying data as well as modify our open-source RAMSES platform to best suit their clinical needs and improve performance.

### A. Methodological limitations

While very encouraging, our study has limitations. One limitation is the relatively small sample size of patient data for the purposes of cross-patient seizure detection. In our cohort, 27 patients have seizures, which likely does not likely contain enough variety to be fully representative of all seizure types and locations encountered in an ICU population. This limitation restricts the types of features and models that we can use to those which perform well on small amounts of data. Furthermore, we do not have training labels for other clinically important phenomena such as sleep stages or different types of non-ictal discharges that may influence classification. Detection of such interictal abnormalities is poor with all commercially available software. Furthermore, the current implementation streams retrospective data from the ieeg.org platform rather than a clinical EEG machine. Based on the results of this pilot study, future work will address this. Despite these shortcomings, we believe that RAMSES is both novel and an important step in moving toward automated monitoring systems that can be implemented over a very short time frame, and dramatically reduce cost and clinician and technologist time in busy cEEG monitoring settings. We believe that focusing on data reduction, rather than just seizure detection is a significant advance that compliments currently available computational EEG methods.

### B. Future directions

The RAMSES system represents a significant starting point for future work in data-driven ICU EEG analysis. We have previously reported an integrated data management and caretaker notification platform for multimodal ICU data[21]. Our vision is for a unifying data platform that is capable of incorporating any number of analytic engines, harnessing the power of cloud computing, and providing real time clinical updates. In the near term, the system should incorporate active learning, in which RAMSES prompts clinicians to label additional segments of EEG when results are equivocal or unsatisfactory. These new markings would then feed into training data and improve the model. Eventually, as additional data streams feed into the system, the data-reduction algorithm could be replaced with a deep-learning model that would require more computational resources than are present in typical hospital computers. Indeed, implantable devices for seizure-detection and stimulation may operate under such a paradigm in the near future[22]. At present, the modular structure of RAMSES ensures that these future iterations can be incorporated without disruption of the system. It is not far-fetched, given our open data and algorithm sharing, to envision this data set dramatically expanding, as we set up multi-institutional collaborations to enhance our performance, data acquisition, and system testing.

Looking forward to implementation, it is interesting to consider what clinical translation pathway might be most expeditious to getting our system integrated into patient care. A typical pathway for these kinds of innovations is to file patents and copyrights, license and co-develop technology with commercial vendors, and then sell closed, propriety systems[23]. This commercial path provides vital accountability for system performance and safety to the FDA, at the cost of slowing development and increased cost[24]. Open source medical software[25], such as OsiriX[26] in radiology and the Veteran's Administration EHR VISTA present interesting models that might be implemented to keep collaboration high and costs low. This type of public-private model is still rare in the field of diagnostics, and presents challenges to our current system of approval, regulatory and legal liability. It may be that our decision to make all of our code and data openly available will advance science in this realm, but potentially delay commercialization of our system, which could delay eventual clinical translational outside of our own institution.

## V. CONCLUSION

As cEEG continues to gain traction at medical centers worldwide, computational tools which can improve the accuracy and speed of evaluation are desperately needed to handle the immense volume of data and the tremendous human and financial costs of manual review. In this study we demonstrated a novel quantitative approach to semi-automated analysis of continuous EEG in the intensive care unit and highlight how it could be deployed to significantly alleviate the burden of manual EEG review while retaining the vast majority of seizures. We share all data and code as well as package our algorithm inside a full-stack application which provides data streaming and a clinician-facing web interface to facilitate clinical translation of our methods. We also present thoughts on the implications of developing medical systems like ours in an open-source framework, and how this might affect eventual commercialization and clinical translation of work like ours. Our ultimate goal is for many more critically-ill patients to benefit from continuous EEG monitoring during their hospitalizations, while reducing the costs and improving the effectiveness of these systems. We present this work as an important step towards automating a task that many experts feel is better suited to machines than manual review.

ACKNOWLEDGMENT

The authors would like to thank Jacqueline A. Boccanfuso and Amanda Samuel for their assistance in uploading data to ieeg.org.

FIGURES AND CAPTIONS

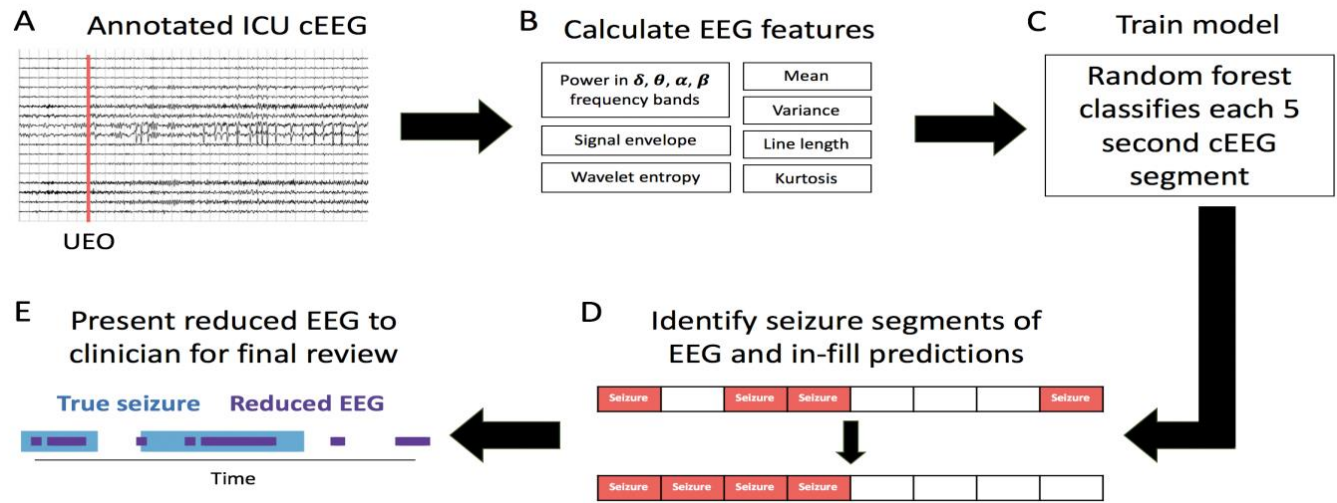

**Fig. 1: Seizure detection & data reduction methods.** (A) we use clinically annotated ICU cEEG (UEO = unequivocal electrographic onset). (B) We calculate the listed EEG features for each channel subtracted from a common average reference before taking the mean and variance of each feature across channels. (C) We train a random forest model to classify each 5 second cEEG segment as likely or unlikely to contain seizure, and test in unseen patients. (D) We smooth predictions to improve interpretability for future clinical review in (E).

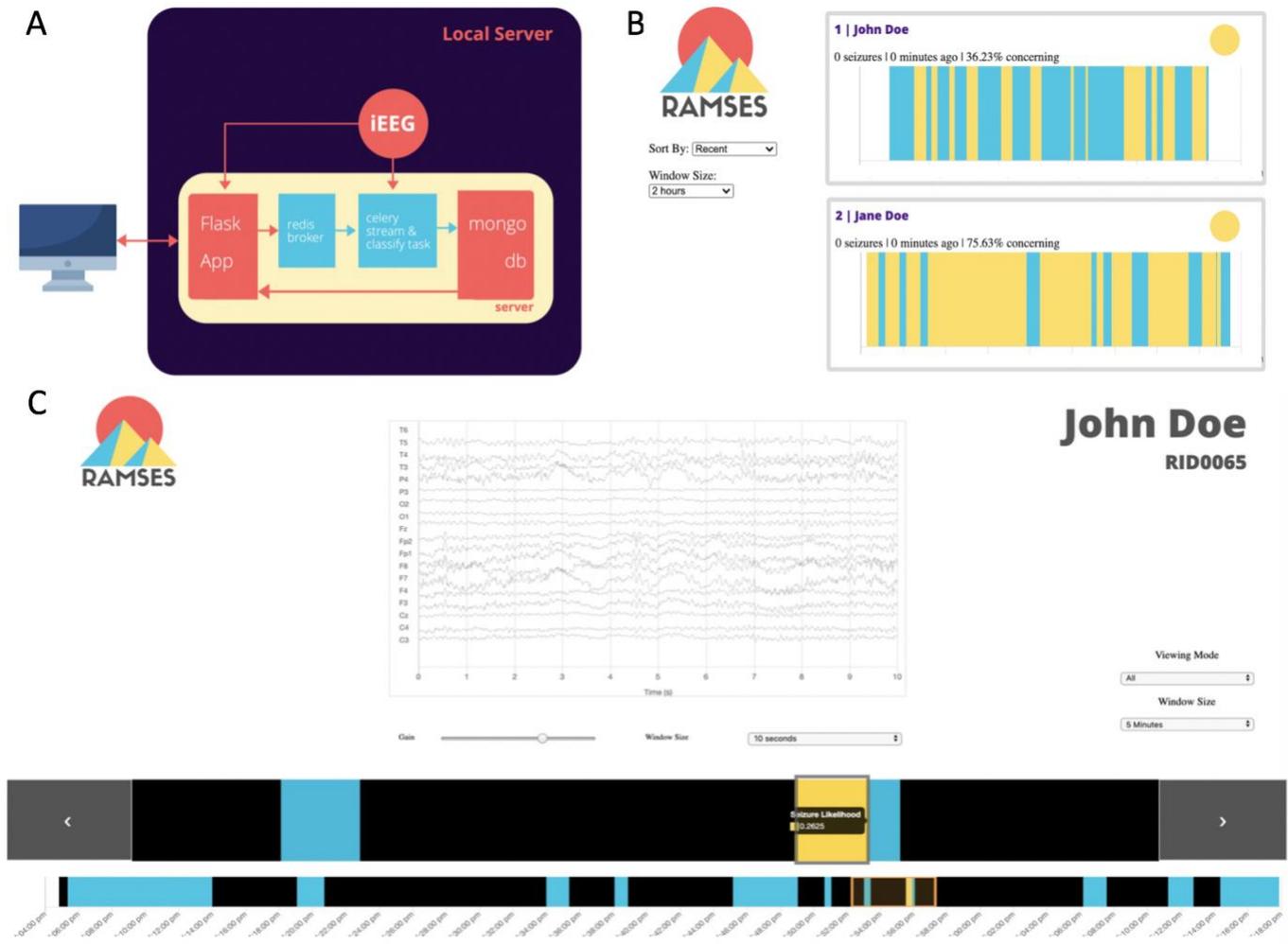

**Fig 2. Real-time alerting and monitoring system for epilepsy and seizures (RAMSES).** (A) Celery workers asynchronously stream data from iEEG.org and store predictions in a mongoDB database which the Flask-based web dashboard queries on user request. (B) Front page of web dashboard showing summary statistics for each patient as well as options to customize the ordering of patients and the length of time over which the statistics are calculated. The dot next to each patient's name indicates the most concerning prediction the algorithm produced over the selected window of time for that patient. (C) Layout of a single patient with raw EEG displayed as well as a navigable timeline of algorithm outputs.

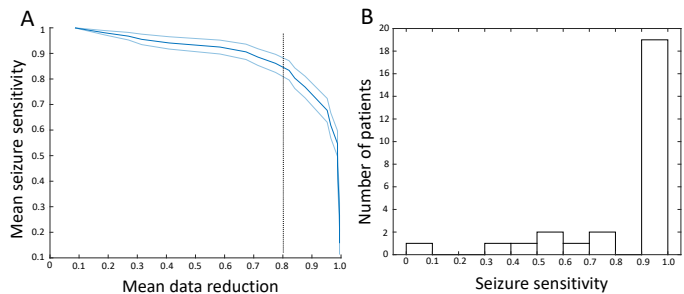

**Fig 3: Algorithm performance.** (A) Tradeoff between seizure sensitivity and mean data reduction. Vertical line: 80% data reduction, blue shading: standard error. (B) Histogram of seizure sensitivities at 80% data reduction dotted vertical line in panel A.

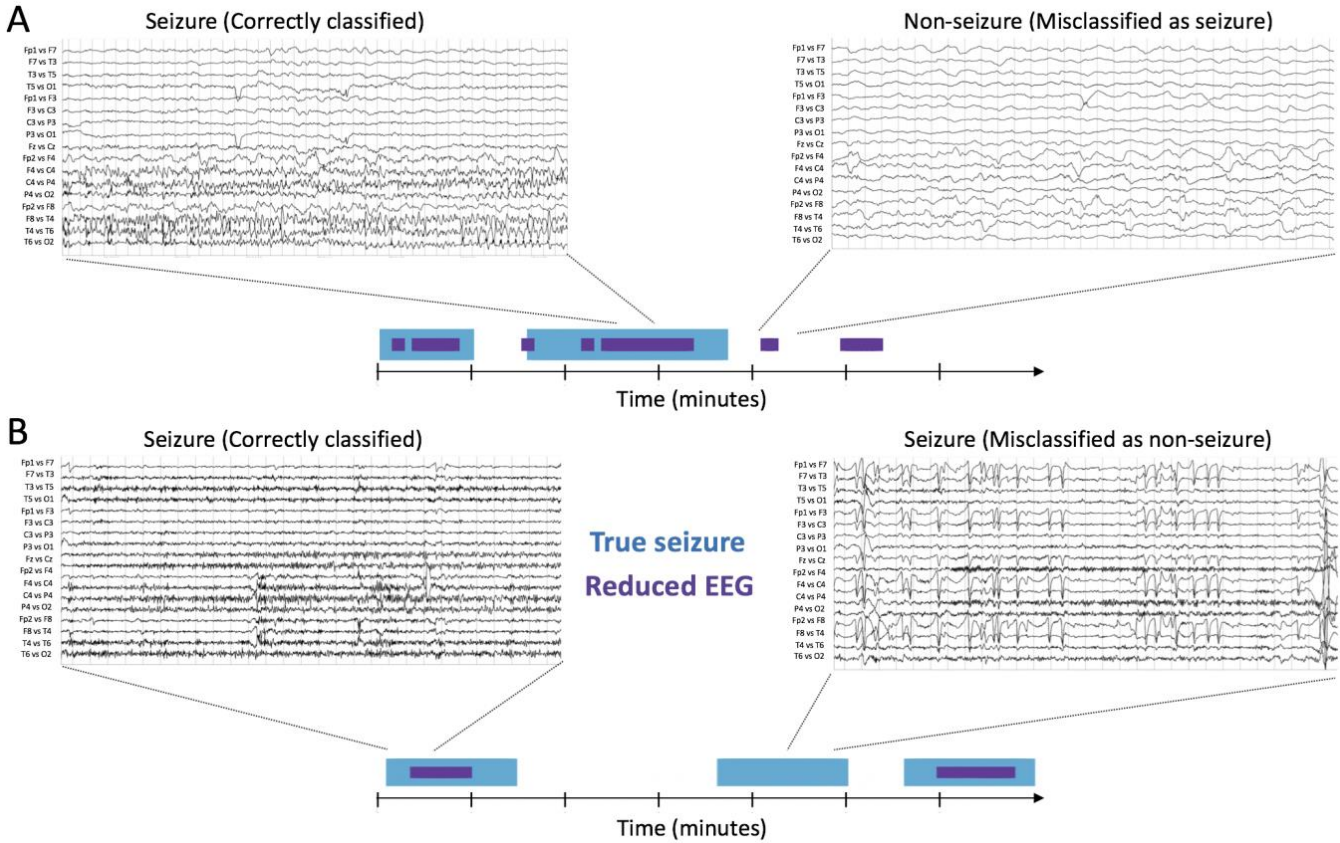

**Fig. 4: Representative results of data reduction algorithm.** In both panels, the distribution of true seizures over an eight-minute period are shown in blue and the reduced EEG is shown in purple. All EEG is displayed in anterior – posterior bipolar montage and is of 35 seconds in length. (A) cEEG clip of a true positive (left) and false positive (right) seizure segments. (B) cEEG clip of true positive (left) and false negative (right) seizure segments.